\documentclass[pre,preprint,showpacs, amssymb, amsmath, preprintnumbers, aps]{revtex4}
\usepackage{graphicx}
\usepackage{amsfonts}
\usepackage{amssymb}
\usepackage{amsmath}

\newcommand{\revision}{}

\begin{document}

% \title{Ion transport in long conical pores: a perturbation
%treatment of the Poisson-Nernst-Planck model.}

\title{Rectification in synthetic conical nanopores: a one-dimensional
Poisson-Nernst-Planck modeling}

\author{I. D. Kosi\'nska}
\affiliation{Institut f\"ur Physik, Universit\"at Augsburg, D-86135 Augsburg,
Germany\\M. Smoluchowski Institute of Physics, Jagiellonian University,
PL-30-059 Krak\'ow, Poland}

 \author{I. Goychuk}
\affiliation{Institut f\"ur Physik, Universit\"at Augsburg, D-86135 Augsburg,
Germany}

 \author{M. Kostur}
\affiliation{Institut f\"ur Physik, Universit\"at Augsburg, D-86135 Augsburg,
Germany}

 \author{G. Schmid}
\affiliation{Institut f\"ur Physik, Universit\"at Augsburg, D-86135 Augsburg,
Germany}

 \author{P. H\"anggi}
\affiliation{Institut f\"ur Physik, Universit\"at Augsburg, D-86135 Augsburg,
Germany\\ Department of Physics, National University of Singapore, Singapore 117542, Republic of Singapore}

\begin{abstract}

  Ion transport in biological and synthetic nanochannels is
  characterized by  phenomena such as ion current fluctuations and
  rectification.  Recently, it has been demonstrated that  nanofabricated
  synthetic pores can mimic  transport properties of biological
  ion channels [P. Yu. Apel, {\it et al.}, Nucl. Instr. Meth. B {\bf 184},
  337 (2001);  Z. Siwy, {\it et al.}, Europhys. Lett. {\bf 60}, 349 (2002)].
  Here, the ion current rectification is studied within a
  reduced 1D Poisson-Nernst-Planck (PNP) model of synthetic nanopores.  A
  conical channel of a few $\mathrm{nm}$ to a few hundred of nm in diameter, and of few
  $\mu$m long is considered in the limit where the channel length considerably
  exceeds the Debye screening length.  The rigid channel wall is
  assumed to be weakly charged. A one-dimensional reduction of the
  three-dimensional problem in terms of corresponding  entropic effects is put forward.
  The ion transport is described by the non-equilibrium steady-state
  solution of the 1D Poisson-Nernst-Planck system within a singular
  perturbation treatment. An analytic formula for the approximate
   rectification current in the lowest order perturbation theory is derived.
  A detailed comparison between  numerical results and
  the singular perturbation theory is presented. The crucial importance of the asymmetry
  in the potential jumps at the pore ends
  on the rectification effect is demonstrated. This so constructed 1D theory is
  shown to describe well the experimental data in the regime of small-to-moderate
  electric currents.

\end{abstract}


\date{\today}

\maketitle

\section{Introduction}

The interaction of a cell with the extracellular environment through
the membrane is of prominent importance for the cell behavior and
cellular function. The lipid bilayer of cell membranes constitutes a
barrier to the passage of charged and polar molecules
\cite{Alberts}.  The channel proteins, which form narrow hydrophilic
pores,  primarily allow for the passage of small inorganic ions. The
opening and closing of ion channels characteristically  depend on
the membrane voltage or the binding of selective ligand molecules
\cite{Hille}.

The production techniques and the application of single nanochannels
in polymer films with lengths of micrometers and diameter sizes  on
the nanoscale  is presently attracting  widespread interest
\cite{Siwy}. Synthetic nanochannels have been fabricated as mimics
of real biological nanopores and ion channels \cite{Siwy2}. A recent
intriguing finding was that conically shaped nanopores can be
put to work as promising sensing elements for small molecules
\cite{Heins}, DNAs \cite{Harrell}, proteins \cite{Siwy3} and yet
other substances \cite{Lee}.

The sensing and transport properties of conical nanopores are
strongly dependent on the shape of the pores, i.e. on
parameters such as  the cone angle and its length.  An
important feature of charged conical nanopores in comparison to
the cylindrical ones is that the voltage drop caused by the ion
current is centered around the narrow tip \cite{Choi}.

The flow of ions through a  nanopore can either  be driven by
concentration gradients or an electric field, or both.  It can
 approximately be modeled  by means of an electro-diffusion equation.
Within this description, the electric field inside the pore is
governed in a self-consistent manner by the ion concentrations via
the Poisson equation. At the boundaries, both the ion (bulk)
concentrations and the electrical potentials are externally fixed.
This may generate a steady current flow. With this work, our
objective is the derivation of the general solution of such a
non-equilibrium steady-state  problem. A challenge is then the
investigation of the influence of an externally applied  voltage on
different polarity for rectification, being induced by the asymmetry
of the nanopore.

A three-dimensional PNP modeling of the ion conductance in biological gramicidin
A channels has been put forward in Ref. \cite{Kurnikova}
for several membrane electrostatics. This 3D
modeling was next investigated in Ref. \cite{Mamonov} in the
context of the so called Potential-of-Mean-Force-Poisson-Nernst-Planck theory which
accounts for the dynamical relaxation of the channel forming protein
and the surrounding medium by incorporating
the free energy of inserting a single ion into the channel.
Furthermore, OmpF and $\alpha$-hemolysin porin channels were studied
within similar approaches, complemented by molecular dynamics, and
Brownian dynamics simulations in Ref. \cite{Im, Noskov},
and standard methods of continuum electrostatics recalled in Ref.~\cite{Alcaraz}. In particular,
for these porin channels, an asymmetric current-voltage characteristics, which
implies a current rectification under alternating voltage conditions, has been observed.
However, the structural complexity of biological ion channels and, in particular,
an inhomogeneous charge distribution on the channel wall prevented any possibility
for a rigorous analytical
treatment in these numerical studies. What is more, it remains
unclear until now how a very popular one-dimensional, textbook \cite{Jackson}
description of the ionic conductance of such and similar biological
channels can be justified from 3D picture if the channel's length
$L$ and diameter $D$ does not differ much, like $L/D\sim 3$ for porins.

Synthetic nanochannels \cite{Siwy}, which have microns in length, are
very much different in this respect providing a nice testbed for
a one-dimensional reduced description.
Such channels are wide enough (up to several nanometers in the narrowest part),
so that the self-consistent mean field PNP theory, which neglects totally
the correlation effects (existing e.g. due to a finite size of ions), seems well applicable.
Moreover, the
conductance of an ion channel in the limit, where its length exceeds
largely the screening Debye length, should be strictly Ohmic \cite{Sneyd},
if the diameter does not vary and the channel wall is not charged.
If the mobilities of free cations and anions are equal,
the rectification effect can therefore only emerge due to a synergy of the entropic
driving force caused by the channel asymmetry (conical pore) and the electrostatic
effects caused by the fixed charges present on the channel wall.
An analytical theory for such a system is currently absent.

With this work we shall address this challenge within a singular perturbation
theory approach for long conical pores, where the pore length
substantially exceeds  both its diameter and the  Debye
length. The obtained analytical results are validated by
precise numerics. Moreover, numerical results are obtained for realistic
surface charge densities beyond the validity range of perturbation theory.
Then, the satisfying agreement with experimental data is
demonstrated.

\section{Ionic transport}

\subsection*{Three dimensional Poisson-Nernst-Planck (PNP) equations}

\begin{figure}
\begin{center}
 \includegraphics[width=8cm]{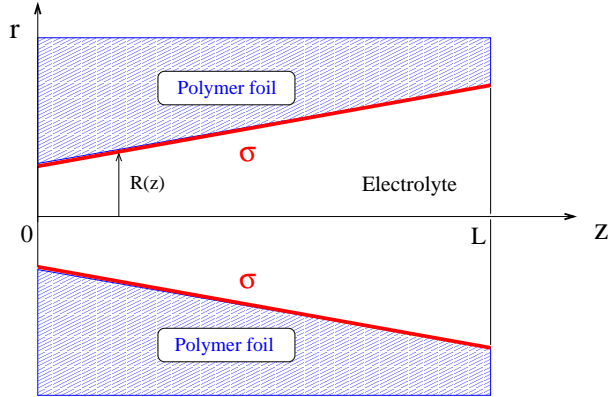}
 \caption{\label{fig1}
(Color online) A sketch of the profile of the conical nanopore. The wall
   inside the pore exhibits a surface charge density $\sigma$. The local radius of the pore,
   $R(z)$, is given by Eq. (\ref{radius}). The drawing
   does not reflect the actual physical ($z:r$)-proportions.}
\end{center}
\end{figure}

We are interested in the ionic currents through relatively narrow
pores which are created in a thin (of $\mu$m) dielectric film.  The
channel wall is assumed to be weakly charged by a prescribed surface
charge density $\sigma$, (see Fig.~\ref{fig1}).
The whole system is immersed in an electrolyte solution. We consider
the limit of long channels with the length being much larger than
both, the pore diameter and the Debye screening length:
$\xi_D=\sqrt{\epsilon_0\epsilon_w k_{\rm{B}} T/(2e^2N_A 10^3 I_c)}$ (in meters).
Therein, $\epsilon_0$ denotes the dielectric constant of vacuum,
$\epsilon_w\approx 80$ the relative dielectric constant of
water, $k_{\rm{B}}$ the Boltzmann constant, $T$ the
temperature, $e$ the elementary charge, and $N_A$ is the Avogadro
number. $I_c=\sum_{i=\pm}\nu_i^2c_i(\infty)/2$ defines the ionic
strength of the electrolyte \cite{Jackson}. It is given in terms of  the bulk
concentration of anions, $c_-(\infty)$, and cations $c_+(\infty)$
(in molars $\equiv$  moles/liter, M), respectively,  with the
valences being $\nu_i$. For monovalent electrolytes i.e.
$\nu_i=\pm 1$, with $c_+(\infty)=c_-(\infty)= 1$ M,
the ionic strength is $I_c=1$
M and thus $\xi_D\approx 0.305$ nm at room temperature $T=298$ K.
This presents a typical experimental situation for the synthetic
pores studied in Ref. \cite{Siwy,Siwy2,Siwy3}, where $\xi_D\sim 1$
nm at sub-molar concentrations of ca. $0.1$ M.
%Moreover, the diameter of the pore $d$ is also wide enough
%to assume Boltzmann distribution of ions in the transverse
%direction i.e $\xi_D \ll d$, where $\xi_D$ is the Debye length.
Our electrolyte system consists of two kinds of ions, potassium, K$^+$, and
chloride, Cl$^{-}$.

Let us next start from the 3D electro-diffusion equation
\begin{equation}
\displaystyle\frac{\partial c_i(\vec{r},t)}{\partial t} = -
\nabla \cdot \vec{j_i}(\vec{r},t)
\label{eq1}
\end{equation}
where
\begin{equation}\label{flux}
\vec{j_i}(\vec{r},t) = -D_i \nabla c_i(\vec{r},t)+
e\nu_i \mu_i \vec{\mathcal{E}}(\vec{r}) c_i(\vec{r},t)
\end{equation}
defines the mass {\revision flux density}
%current
$\vec{j_i}$ of the $i$-th species  of ions,
$\vec{r}$ is the position vector, $D_i$ the diffusion
constant,  $\mu_i$ the mobility of the ion particles, and
$\vec{\mathcal{E}}$ denotes the electric field. The consistency with
the thermal equilibrium demands that the mobility of the ions
$\mu_i$ fulfills the Sutherland-Einstein relation, reading
$D_i=\mu_i/\beta$, with the inverse temperature $\beta=1/k_{\rm{B}} T$.
Because $\vec{\mathcal{E}}(\vec{r})=-\nabla \Phi(\vec{r})$,
where $\Phi(\vec{r})$ is the electric potential, one can recast
 Eq. (\ref{eq1}) into the  more convenient form:
\begin{equation}
\displaystyle\frac{\partial c_i(\vec{r},t)}{\partial t}=
D_i\, {\Huge \nabla} \, \exp[-e\nu_i\, \beta\, \Phi(\vec{r})]\, {\huge \nabla} \,
\exp[\,e\nu_i \, \beta \, \Phi(\vec{r})]\,c_i(\vec{r},t).
\label{smol}
\end{equation}
The appropriate choice of coordinates for the problem at hand are
the cylindrical coordinates $(r,\phi,z)$. The electric potential
$\Phi(r,z)$ is governed in a self-consistent manner by the Poisson
equation
\begin{equation}
\epsilon_0\nabla\cdot[\epsilon(r,z)\nabla\Phi(r,z)] = -\sum_i \rho_i(r,z)  - \rho_{\rm{fix}}(r,z).
\label{Poisson}
\end{equation}
The ion
concentration $c_i$ (in molars) is related to the density of
electric charge $\rho_i$ (in units of Coulombs per cubic meter, C/m$^{3}$) through
$\rho_i = 10^3  \; e\nu_i \; N_A \; \Theta\left(-r+R(z)\right) \; c_i$,
where $\Theta$ is the Heaviside step function, and
\begin{equation}
R(z) = R(0) + (R(L) - R(0))z/L
\label{radius}
\end{equation}
denotes a variable cone radius.
The nanochannel
radii at the left and right pore ends are $R(0)$ and $R(L)$,
respectively, and $L$ is the length of the pore.
 The charge density $\rho_{\rm{fix}} = \delta(r-R(z))\sigma$, with $\delta(x)$ being the Dirac delta-function,
 represents the fixed
charges located on the inside of the channel wall. $\epsilon(r,z) =
\epsilon_p \Theta(r-R(z)) + \epsilon_w \Theta(-r+R(z))$ describes
dielectric properties of the open nanopore, where $\epsilon_p,
\epsilon_w$ are the relative dielectric constants of the polymer and
water, respectively. The set  of relations (\ref{smol}) and
(\ref{Poisson}) for $\nabla \cdot \vec{j_i}=0$ constitutes the system of
coupled Poisson-Nernst-Planck (PNP) equations.

\section{One-dimensional model reduction for long pores}

{\revision The primary purpose of our work is to gain  analytical insight
into the nature of rectification current.} Since an analytical
solution for the 3D PNP in general could not be found, a
simplification is  unavoidable. The ion diffusion and the ion flow are
confined by the channel to a variable cross-section area $\pi
R^2(z)$. Assuming local equilibrium in the transverse ($r$)
direction, the full three dimensional system can be cast into an
effective one dimensional problem. The price to be paid is the
emergence of an additional, entropic potential contribution $\Delta
S(z)$ \cite{Zwanzig, Rubi,Reguera1,Reguera2}. {\revision Moreover, as
pointed out by Zwanzig \cite{Zwanzig}, a renormalization of the
diffusion coefficient emerges as a dynamical effect  due to a finite
relaxation time in the transversal direction. In our situation
(i.e. for a small opening angle of the cone) this latter dynamical correction
to the diffusion coefficient is negligible \cite{Bezrukov}. If the resulting currents
are too large these model assumptions can not be justified, 
as the particles must have sufficient time
to relax into the transverse direction \cite{Reguera2}.}

%and a renormalization of the
%diffusion coefficient, being due to a finite, non-zero relaxation
%time in the transversal direction. For the parameters considered in
%this paper, the latter effect is very small and thus can safely  be
%neglected.

In  cylindrical coordinates $(r,\phi,z)$ the equation for mass
current due to the coaxial symmetry assumes the form
%\begin{equation}
\begin{align}
\vec{j_i}(r, z,t) =& -D_i\exp[-e\nu_i \, \beta \, \Phi(r,z)]
\,\frac{\partial}{\partial r} \exp[\, e\nu_i \, \beta \, \Phi(r,z)]\,
c_i(r,z,t)\,\hat{r} \nonumber \\
&-D_i\exp[-e\nu_i \, \beta \, \Phi(r,z)]\,\frac{\partial}{\partial z}
\exp[\, e\nu_i \, \beta \, \Phi(r,z)]\,c_i(r,z,t)\,\hat{z},
\end{align}
%\end{equation}
where $\hat{r}$ and $\hat{z}$ are the units vectors in cylindrical
coordinates.

%Our system consists of two kinds of ions, K$^+$ and Cl$^{-}$.  The
%concentration $c_i$ with $i=\rm{K}^+,~\rm{Cl}^-$ for each given type
%of ions, should fulfil Eq. (\ref{eq1}).
The steady-state regime is described by the condition:
\begin{equation}
\nabla\cdot \vec{j_i} = 0.
\end{equation}

Because the length of the channel by far  exceeds  its diameter, we
can assume local equilibrium, i.e. uniform distributions for the ions and a constant
potential across each channel cross-section,
 \begin{equation}
\begin{array}{cc}
c(z,r,t) \approx c(z,t),~~ & \Phi(r,z,t) \approx \Phi(z,t).
%\vec{j}_i = j_i \hat{z}.
\label{constant}
\end{array}
\end{equation}
The validity of  an instant local equilibrium in
 transversal direction is fulfilled for weakly charged
channel walls and not too large ion flow through the channel.
Under this assumption, it is possible to reduce the complex
original 3D problem to a simpler, effective  one-dimensional
description. \\
%This approximation
% means that the pore is considered
%like a metal in the transverse direction with infinitely mobile ions.

Using the divergence of a given vector $\vec{u}$
\begin{equation}
\nabla\cdot \vec{u} = \lim_{\Delta z\to 0}\frac{1}{A(z)\Delta z}\oint_{S} \vec{u}\cdot \hat{n}~dS,
\label{div}
\end{equation}
where $V(z)=A(z)\Delta z$ denotes the volume surrounded by the
closed surface $S$ with the normal vector $\hat{n}$, and the fact that no ion flux across the channel
wall occurs,  Eq. (\ref{eq1}) can be reduced \cite{Gardner} under the
assumptions in Eq. (\ref{constant}) to the form
\begin{equation}
%\begin{multline}
\frac{\partial \bar{c}_i(z,t)}{\partial t} =
\frac{\partial}{\partial z}\bigg\{\nu_i e \, \beta \, D_i \, \bar{c}_i(z,t)
\frac{\partial \Phi(z)}{\partial z}
+ A(z) \, D_i \, \frac{\partial}{\partial z}\frac{\bar{c}_i(z,t)}{A(z)}  \bigg\},
%-j_{i}/D_i = \frac{\partial}{\partial z} c_i(r,z) + \nu_i \beta e c_i(r,z)\frac{\partial}{\partial z}\Phi(r,z),
\label{smol_1}
%\end{multline}
\end{equation}
wherein $\bar{c}_i(z,t) = 10^3 \, c_i(z,t)A(z)$ denote the one-dimensional
concentrations (in moles per meter).

Likewise, using  (\ref{div}) and the electrostatic boundary condition
on the channel wall, reading
\begin{equation}
\epsilon_{p}\mathcal{E}^{\perp}_{\rm{polymer}} -
\epsilon_{w}\mathcal{E}^{\perp} = \sigma,
\end{equation}
%where $-\sigma$ denotes a negative surface-charge density,
the 3D Poisson equation is effectively reduced \cite{Gillespie,Gardner} into the form
\begin{align}
\displaystyle \epsilon_0\epsilon_w\frac{d^2 \Phi(z)}{d z^2} =
-\frac{2\sigma}{R(z)} - e N_A \sum_i\nu_i \frac{\bar{c}_i(z,t)}{A(z)}
- \epsilon_0\epsilon_w\frac{d\Phi(z)}{dz}\frac{d}{dz}\ln A(z).
\label{Poisson_mod}
\end{align}
The quantity $2\sigma/R(z)$ has the meaning of the volume density of the fixed surface charge,
$\rho = \Delta Q/\Delta V$, where $\Delta Q = 2 \pi \sigma R(z) \Delta z$ and $\Delta V = \pi R^2(z) \Delta z$.

Likewise, Eq. (\ref{smol_1}) can be recast as
\begin{equation}\label{FJ}
\frac{\partial \bar{c}_i(z,t)}{\partial t} = -\frac{\partial}{\partial z} J_{i}(z,t) =
\frac{\partial}{\partial z}\,\bigg\{D_i \, \exp[-\beta G_i(z)] \,
\frac{\partial}{\partial z}\, \bar{c}_i(z,t) \, \exp[\,\beta G_i(z)]\bigg\},
\end{equation}
where {\revision $J_i$ denotes mass fluxes, and }$G_i(z)$ is the free energy (potential of mean force), given
by
\begin{equation}
G_i(z) = -T S(z) + e\nu_i \; \Phi(z).
\end{equation}
Due to the reduction from 3D into 1D it inherits the entropic
contribution
\begin{equation}\label{entropic}
\displaystyle S(z) = k_{\rm{B}}\ln \left( A(z)/A_0\right),
\end{equation}
 where $A_0$ denotes an arbitrary, but irrelevant reference cross-section area.  The
coupled nonlinear equations (\ref{Poisson_mod}) together with the
set of Eqs. (\ref{FJ})-(\ref{entropic}) present a self-consistent,
reduced 1D electro-diffusion equation which is similar in
spirit to the well-known  reduction of a 3D-diffusion problem to an
effective Fick-Jacobs diffusion equation \cite{Zwanzig,Rubi,
Reguera1,Reguera2}. The dependence $R(z)$ is still arbitrary within
the discussed approximations; its variation must be smooth, however,
such that the $z$-variations of the normal to the channel's surface
from the cylindrical geometry become negligible up to the second
order. This constructed  1D Poisson-Nernst-Planck modeling
generalizes the one given in  Ref. \cite{Gardner} for  strictly
cylindrical pores to nanopores possessing a  variable cross-section
diameter.

\subsection{Evaluation of the electric currents}

We consider the steady-state, non-equilibrium solutions
characterized by {\revision constant mass fluxes $J_i$. The corresponding
electric currents are $I_i=F\nu_i J_i$,} where $F$ denotes the
Faraday constant. Explicitly, these electric currents for the $i-th$
ion species read:
\begin{equation}
I_i= - F \, D_i \, \nu_i \exp[-\beta G_i(z)]\,
\frac{d}{d z}\, \bar{c}_i(z) \, \exp[\,\beta G_i(z)]\;.
\label{flux_steady}
\end{equation}
Thus,  currents are fully determined by the bulk ion-concentrations
and the electric potential difference $U=\Phi(0)-\Phi(L)$ across the
membrane, as determined by the effective 1D-Poisson-Nernst-Planck
equations.

%In 1D form
%\begin{equation}
%\frac{\partial}{\partial z}\bigg( A(z) \epsilon_0\epsilon_w \frac{\partial \Phi}{\partial z}\bigg) =
%2\sigma(z)\pi R(z) - e N_A 10^3\sum_i \nu_i\bar{c}_i
%\end{equation}

\subsection{Dimensionless equations}

For the sake of convenience we  perform all our  numerical and
analytical calculations in dimensionless variables. In doing so, we
transform the Poisson-Nernst-Planck equations to their dimensionless
form by use of following relations: $\displaystyle z^{\star}=z/L$,
$\displaystyle R^{\star}(z)=R(z)/R(0)$, $\Phi^\star(z)=\beta e\:
\Phi(z)$, $I^\star_i = I_i L/(FD_ic_0)$, $\displaystyle
\bar{c}^\star_i (z)= \bar{c_i} (z)/c_0$, where $c_0$ is a reference
1D-con\-cen\-tra\-tion (in $1/6.023\times 10^{-14}$  mole/m).
Then, the equation for the constant
electric currents assume the appealing form
\begin{equation}
\nu_i I^\star_i + \frac{d}{dz^\star}\bar{c}^\star_i + \nu_i\bar{c}^\star_i\frac{d\Phi^\star}{dz^\star} -
\bar{c}^\star_i\frac{2}{R^\star}\frac{d R^\star}{dz^\star} = 0,
\label{current_dim}
\end{equation}
and  (\ref{Poisson_mod}) transforms into:
\begin{equation}
%\begin{multline}
\displaystyle
\frac{1}{\lambda^2}\left( \frac{d^2 \Phi^\star}{d z^{\star 2}} +
\frac{2}{R^\star}\frac{d R^\star}{dz^\star}
\frac{d\Phi^\star}{dz^\star}\right)
 + \frac{\epsilon}{R^\star}  +  \frac{1}{\pi R^{\star 2}} \sum_i \nu_i \bar{c}^\star_i = 0,
\label{Poisson_dim}
%\end{multline}
\end{equation}
with
\begin{align}
1/\lambda^2 &= \displaystyle\frac{\epsilon_0\epsilon_w \, R^2(0)}{c_0 F \: \beta  e\; L^2},
\label{lambda}\\
\intertext{and the effective dimensionless surface charge density}
\displaystyle \epsilon &= \frac{2\sigma R(0)}{c_0 F}.
\label{epsilon}
\end{align}
Below, upon simplifying the notations, we shall  suppress
the notation with the superscript~($^*$).

\subsection{Boundary conditions}

We next solve the system of equations (\ref{current_dim}) and
(\ref{Poisson_dim}) with the following boundary conditions:
\begin{equation}
\begin{array}{cc}
\vspace{10pt}
\bar{c}_{\mathrm{K^+}}(0) = \pi c_{\mathrm{K^+},L},& \bar{c}_{\mathrm{Cl^-}}(0) = \pi c_{\mathrm{Cl^-},L},\\
\vspace{10pt}
\bar{c}_{\mathrm{K^+}}(1) = \pi(1+\gamma)^2 c_{\mathrm{K^+},R},& \bar{c}_{\mathrm{Cl^-}}(1) = \pi (1+\gamma)^2 c_{\mathrm{Cl^-},R},\\
\Phi(0) = 0,& \Phi(1)=\Phi_R. \label{boundary}
\end{array}
\end{equation}
wherein $\gamma=(R(L)-R(z=0))/R(z=0) = R(1) - 1$, $c_{i,\{L,R\}} = 10^3 \, c_{\mathrm{bulk},\{L,R\}}\,
R^2(z=0)/c_0 $ and $c_{\mathrm{bulk}, \{L, R\}}$ denote the bulk
concentrations of ions on the left and right sides, respectively.
The electro-neutrality condition yields:
$c_{\mathrm{K^+},L}=c_{\mathrm{Cl^-},L}=c_{L}$, and $c_{\mathrm{K^+},R}=c_{\mathrm{Cl^-},R}=c_{R}$.
Furthermore, the difference of dimensionless potentials across the
nanopore is related to the applied voltage $U$ (in units of  Volts)
by $\Phi(0)-\Phi(1) = \beta e U$, yielding $U = - \Phi(1) / (\beta e)$.

\section{Singular perturbation theory}

The corresponding 1D system  contains two small parameters, namely
\begin{description}
\item[(i)] $1/\lambda$, in Eq.~(\ref{lambda}), which is proportional to the ratio of the Debye length $\xi_D$
to the channel length $L$
\item[(ii)] the  scaled surface charge
density $\epsilon$ [see Eq.~(\ref{epsilon}) and Appendix].
\end{description}
In our case i.e. a long nanopore
with a typical   length $L = 12000$ nm, $1/\lambda$ is of the order
of $10^{-5}$ for sub-molar concentrations. The parameter $\epsilon$
is of the order of $10^{-1}$, or larger. Therefore
we will focus on the leading term in the
limit $1/\lambda\to 0$, while $\epsilon$ will be considered as a
regular series expansion parameter. Thus, the problem requires to use
a singular perturbation theory in $1/\lambda$ and a regular expansion in $\epsilon$.
We use the standard method of matched asymptotic expansions \cite{Nayfeh}
% applying it to a new 1D PNP problem that
%includes the entropic effects (within a potential of mean force) and a
%modified Poisson equation. Moreover, we compare the analytical results
%obtained with the numerical solutions of our novel 1D PNP equations.
%Assuming that $1/\lambda$ is a small parameter, $\lambda\to \infty$,
and seek for an approximate solution of Eq.(\ref{current_dim}) and
(\ref{Poisson_dim}) in the form
\begin{equation}
\begin{array}{c}
\vspace{10pt}
\displaystyle\Phi = \Phi^{(0)} + \frac{1}{\lambda}\Phi^{(1)}+\dots,\\
\vspace{10pt}
\displaystyle \bar{c}_{\mathrm{K^+}} = c_{\mathrm{K^+}}^{(0)} + \frac{1}{\lambda} c_{\mathrm{K^+}}^{(1)}+\dots,\\
\displaystyle \bar{c}_{\mathrm{Cl^-}} = c_{\mathrm{Cl^-}}^{(0)} + \frac{1}{\lambda} c_{\mathrm{Cl^-}}^{(1)}+\dots.
\end{array}
\end{equation}
Since the small parameter $1/\lambda^2$ appears in Eq.
(\ref{Poisson_dim}) in front of  the derivatives, we are dealing
with a typical singularly perturbed boundary-value problem.
%possessing two independent perturbation parameters $1/\lambda$ and
%$\epsilon$. The solution is sought in the leading order of
%$1/\lambda\to 0$, while $\epsilon$ is considered as the expansion
%parameter (see Appendix).

Below we consider a symmetric bulk situation
with  equal ion concentrations on both membrane sides,
$c_{L} = c_{R} = c$. After  cumbersome calculations the uniformly
valid approximation on $[0, 1]$  is found in the first order of
$\epsilon$ to read:

\begin{equation}
%\begin{array}{c}
%\vspace{10pt}
%\begin{align}
\Phi(z) \approx \Phi^{(0)}(z)=  \Phi_{0}(z) -
\epsilon
\bigg\{
\phi_{1}^L(\lambda z) +
\Phi_{1}(z)  +
\phi^R_{1}(\lambda (1-z))
 + \displaystyle\frac{1}{2c}+
\displaystyle
\frac{1}{2c(1+\gamma)}\bigg\} + O(\epsilon^2),
\label{pot_uni}
%\end{align}
%\end{array}
\end{equation}
%\begin{equation}
%\begin{array}{c}
%\vspace{10pt}
\begin{multline}
\displaystyle \bar{c}_{\{\mathrm{K^+, Cl^-}\}}(z) = \frac{1}{2}
c_{\Sigma, 0}(z)
- \frac{1}{2} \epsilon\bigg\{ \pm\pi/R(z)
+ c_{\Sigma, 1}(z) \\ \mp
2\pi c\phi^L_{1}(\lambda z)
\mp 2\pi c (1+\gamma)^2 \;
\phi^R_{1}(\lambda (1-z)) \mp \pi(2+\gamma) \bigg\}
 + O(\epsilon^2),
\label{con_pos_uni}
\end{multline}
%\end{array}
%\end{equation}
%{}
%\begin{equation}
%\begin{array}{c}
%\vspace{10pt}
%\displaystyle \bar{c}_{Cl^-}(z) = \frac{1}{2}
%\bigg( c_{K^+}(z) + c_{Cl^-}(z) \bigg)_{(0)} +
%\frac{1}{2} \epsilon\bigg\{\bigg[ -\pi\sigma(z) \\
%+ \bigg(c_{K^+}(z)  +
%c_{Cl^-}(z)\bigg)_{(1)}\bigg] +
%2\pi c\phi_{(1)}(\lambda z) \\
%+ 2\pi c (1+\gamma)^2
%\tilde{\phi}_{(1)}(\lambda (1-z)) + \pi(2+\gamma) \bigg\}
% + O(\epsilon^2),
%\label{con_neg_uni}
%\end{array}
%\end{equation}
where the upper sign refers to $\mathrm{K^+}$
ions, and the lower one to $\mathrm{Cl^-}$, respectively, and
the expressions
%\begin{equation}
%\begin{array}{c}
%\vspace{10pt}
\begin{eqnarray}
 c_{\Sigma, 0}(z) & = &
\left(\frac{-J_{0}}{\gamma\,R(z)} + C_{0}\right)R^{2}(z) = C_0R^2(z),\\ \nonumber
\Phi_{0}(z) & = & \frac{ I_{0}}{\gamma C_{0}}\frac{1}{R(z)} +
 E_{0}  = \Phi_R\frac{(1+\gamma)z}{R(z)}, \\ \nonumber
 c_{\Sigma, 1}(z) & = &
\bigg(
 -\frac{\pi I_{0}}{2\gamma C_{0}}
\frac{1}{R^{2}(z)}
- \frac{J_{1}}{\gamma\, R(z)}
 + C_{1}
\bigg) R^{2} (z) \\ \nonumber
& = & \frac{1}{2}\pi\Phi_R\gamma(z-1)z,
\\ \nonumber
 \Phi_{1}(z) & = &\frac{I_{0}}{C^{2}_{0}
\gamma}
\left(
\frac{\pi I_{0}}{6\gamma C_{0}}\frac{1}{R^{3}(z)}+
\frac{J_{1}}{2\gamma}\frac{1}{R^{2}(z)}
-
\frac{C_{1}}{R(z)}
\right) \\ \nonumber
& + & \frac{I_{1} - \pi\gamma}{C_{0}\gamma \, R(z)} + E_{1},
\end{eqnarray}
%\end{array}
%\end{equation}
contain eight constants $C_0, C_1, I_0, I_1, J_0, J_1, E_0, E_1$ which are determined from the matching
conditions (see in the Appendix).
Furthermore,
the left/right boundary layer potential variations in
Eqs. (\ref{pot_uni}), (\ref{con_pos_uni})
are given, respectively, by
\begin{eqnarray}
\vspace{10pt} \displaystyle\phi_{1}^L(\lambda z) & = &
\frac{1}{2c}\left( \exp[-\sqrt{2 \, c}\lambda z] - 1  \right),\\
\displaystyle \phi_{1}^R(\lambda (1-z)) & = &
\frac{1}{2c(1+\gamma)}\left(\exp[-\sqrt{2\,c}\lambda (1-z)] - 1 \right),
\nonumber
\label{bl}
\end{eqnarray}
where $\sqrt{2c}\lambda/L = 1/\xi_{\rm{D}}$, with $\xi_{\rm{D}}$ being the
Debye length in bulk.

Assuming that
the both sorts of ions have equal diffusion coefficients
\cite{book}, i.e. $D_{\mathrm{K^+}}=D_{\mathrm{Cl^-}} = D$, the two constants $J = - I_{\mathrm{K^+}} + I_{\mathrm{Cl^-}}$ and
$I = I_{\mathrm{K^+}} + I_{\mathrm{Cl^-}}$ (see Ref.~\cite{Barcilon}) have the following physical meaning: $J=J_{0} -
\epsilon J_{1}+ O(\epsilon^2)$ yields an approximation
to the negative of the total mass
current, whereas $I = I_{0}-\epsilon
I_{1}+ O(\epsilon^2)$ approximates the total electric current,
\begin{eqnarray}\label{mres1}
I  \approx - 2\pi c(1+\gamma)\Phi_R\left(1 + \epsilon\frac{1}{24 c}
  \frac{\gamma}{1+\gamma}\Phi_R \right)\;.
\label{i_total}
\end{eqnarray}
Eq.~(\ref{i_total}) presents a main analytical result of this paper.
In the original
physical variables it reads
\begin{eqnarray}
I & = &I_{\mathrm{K^+}}+I_{\mathrm{Cl^-}}\approx
  2\pi F D c_{\mathrm{bulk}} \left(R(0)R(L)/L\right) \left(eU/k_{\rm{B}}T\right)
  \nonumber \\
   &\times & \left[1 - \frac{1}{12}\frac{\sigma}{c_{\mathrm{bulk}}F}
  \left(\frac{1}{R(0)}-\frac{1}{R(L)} \right)\frac{eU}{k_{\rm{B}}T} \right]
\label{cur_tot_first}
\end{eqnarray}
where $c_{\mathrm{bulk}}$ is the concentration of ions in bulk.

It shows that the rectification effect appears already in the first
order of $\epsilon$. The effect vanishes, when the pore is not
charged, $\sigma=0$. This is because the rectification effects for
the monovalent cations and anions for an uncharged pore are
(trivially) ``counter-directed'' and compensate each other exactly
if both sort of ions have equal diffusion coefficients.
Obviously, for the cylindrical pore with $\gamma=0$, it goes away as
well. Apparently, the current rectification is thus due to a synergy
of the entropic effect caused by the pore asymmetry and the surface
charge present. Moreover, one can attribute this result also to different 
volume charge densities $2\sigma/R(z)$ of the fixed charges at the both channels ends. 
Given this latter interpretation, one can expect also rectification effect for cylindrical
pores inhomogeneously charged with two very different charge densities at the ends.

Using the computer algebra
system MAPLE, we obtained also analytical results for
electric currents and fluxes ({\it total} and individual) up to fourth order in $\epsilon$.
The results
are, however, rather cumbersome and of limited analytical insight,
and thus are not explicitly displayed here. We evaluated them
though numerically, see the corresponding figures given below.
Furthermore, the PNP-system in Eqs. (\ref{current_dim}),
(\ref{Poisson_dim}) with the boundary conditions in Eq.
(\ref{boundary}) was integrated numerically
by making use of a collocating method with adaptive meshing \cite{NAG}.
The results
of the analytical perturbation theory and the numerical solutions
compare very favorably where the perturbation theory is expected to
work properly.

\section{Results and numerical comparison}

\subsection{Perturbation theory vs. numerics}

We consider channels with some fixed left opening radius $R(z=0)=3$
nm and two different right opening radii:
\begin{itemize}
\item [-]``long pore'':  ~$L = 12000$ nm \; , $R(L)=220$ nm
\item [-]``short pore'':  $ L = 200$ nm \;\;\;\;\;, $R(L)=6.616$ nm .
\end{itemize}
In both cases the opening angle of the cone is identical, $\psi =
2\arctan[(R(L)-R(z=0))/L]\approx 2^o$. Furthermore, two different
surface charge densities are used in our studies, namely:
$\sigma=-0.02,-0.1$ e/nm$^2$. The first one corresponds to a
parameter value $\epsilon = 0.12$ which is well within the regime of
validity of the perturbation scheme. For the second value we have
$\epsilon = 0.6$ for which we already expect the perturbation method
to fail, but it still might work occasionally. Furthermore, the
diffusion coefficients were taken as
$D_{\mathrm{K^+}}=D_{\mathrm{Cl^-}} = 2\times 10^{9}$ nm$^2$/s, see
in \cite{book}, temperature $T=298$ K and $\epsilon_w=80$.
\begin{figure}[htb!]
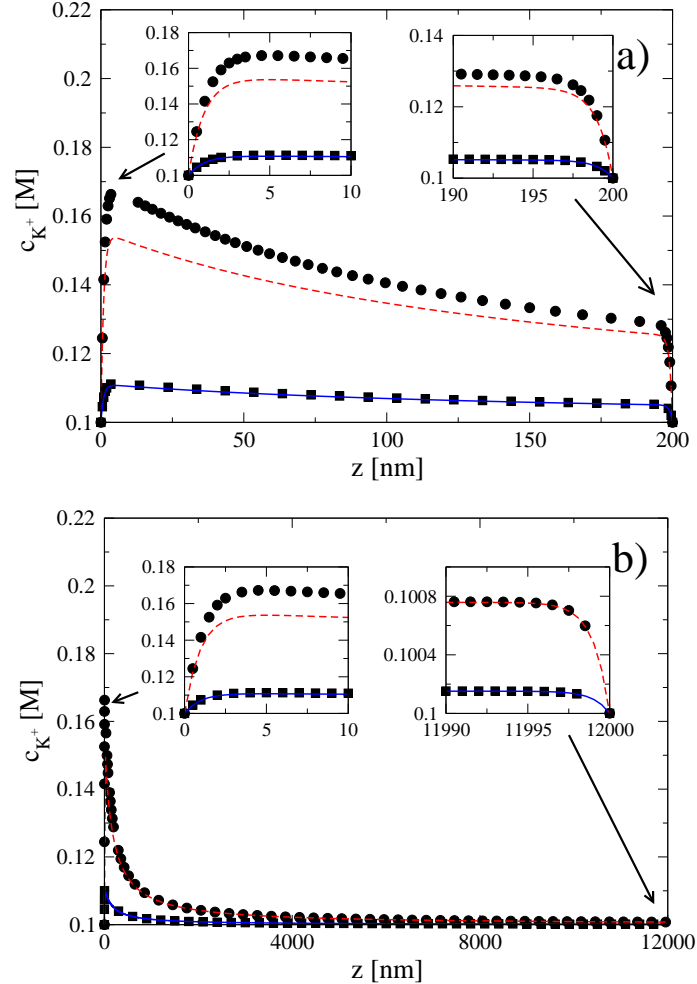

\begin{center}
\includegraphics[width=9cm]{con_pos_k}
\vspace{.3cm}

\includegraphics[width=9cm]{con_pos_l}
 \caption{\label{fig2}
(Color online) Concentration profile $c_{\mathrm{K^+}}(z)$ at
   equilibrium ($c_L=c_R=0.1$ M, $\Phi(z=0)=\Phi(z=L)=0$). Calculations
   are done for the ``short pore'' (a) and the ``long pore'' (b).
   Solid line and squares:
   $\sigma=-0.02$ e/nm$^2$, dashed line and circles:
   $\sigma=-0.1$ e/nm$^2$. Symbols in all cases denote the results of the  numerical
   solution, lines represent the results of the perturbation theory.  The insets depict
   a closer look into the left and the right boundary layers, respectively.}
\end{center}
\end{figure}
The concentration profiles for both, the singular perturbation
solution and the numerical solution are depicted in Fig.~\ref{fig2}.
The first order approximation agrees very well with numerical
solution for the small surface charge density of $\sigma=-0.02$
e/nm$^2$. However, this agreement is worsening upon increasing the
values of $\sigma$. For the  moderately large charge density
$\sigma=-0.1$ e/nm$^2$ the discrepancy between the exact solution
and analytical approximation becomes already significant. In the
inset in Fig.~\ref{fig2} we show the left and the right boundary
layers, according to Eq.~(\ref{con_pos_uni}). Both layers possess a width of several Debye lengths
$\xi_D$. In the case of the
``long pore'', the increase of the concentration $c_{\mathrm{K^+}}(z)$ within the
boundary layer is more  distinct near the narrow opening (the
boundary $z = 0$), as compared to the wide opening at the right
boundary at $z=L$, see Fig.~\ref{fig2}b. This is so because for this
channel the right opening radius  $R(L)=220$ nm is very wide. The
shorter channel with left and right opening radii of  similar size
does not display such a striking difference.

\subsection{Current-Voltage characteristics and rectification}

\begin{figure}[htb!]
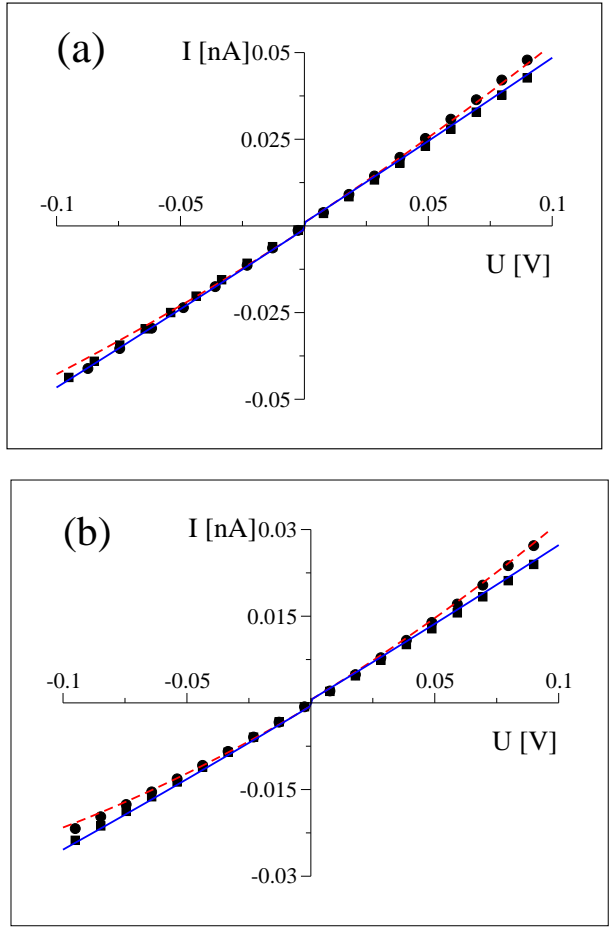

\begin{center}
 \includegraphics[width=8.0cm]{prad_k_bis}
\vspace{.3cm}

 \includegraphics[width=8.0cm]{prad_l_bis}
 \caption{\label{fig3a}
(Color online) Current voltage ($I-U$) characteristics of the
   ``short pore'' (a) and the ``long pore'' (b) for $U$ from $-0.1$ V to $0.1$ V.
   Solid line and squares:
   $\sigma=-0.02$ e/nm$^2$, dashed line and circles:
   $\sigma=-0.1$ e/nm$^2$. Symbols denote the numerical
   solution, the lines represent the results of first order perturbation theory (in $\epsilon$).}
\end{center}
\end{figure}

\begin{figure}[htb!]
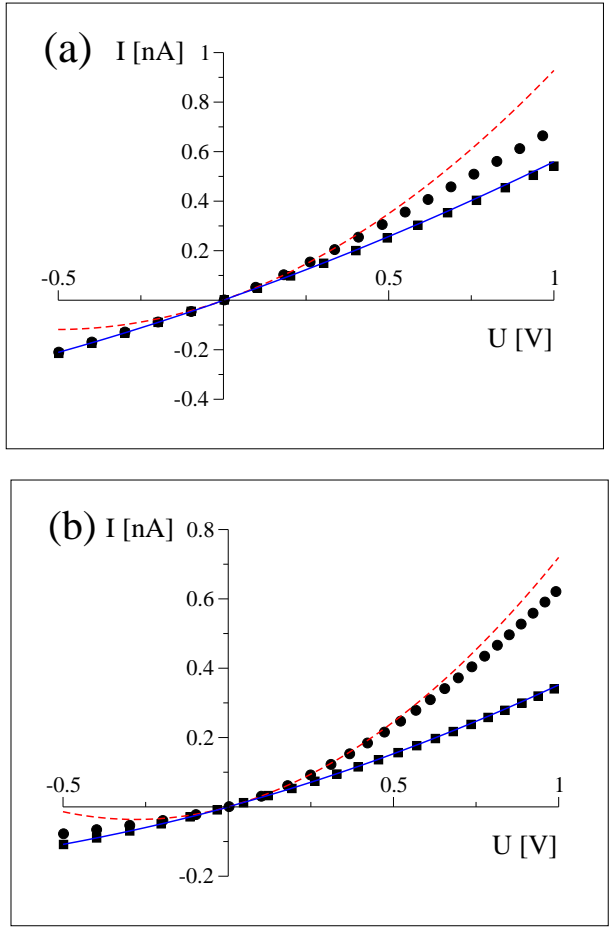

\begin{center}
 \includegraphics[width=8.0cm]{prad_k}
\vspace{.3cm}

 \includegraphics[width=8.0cm]{prad_l}
 \caption{\label{fig3}
(Color online) Current voltage ($I-U$) characteristics of the
   ``short pore'' (a) and the ``long pore'' (b) for $U$ from $-0.5$ V to $1$ V.
   Solid line and squares:
   $\sigma=-0.02$ e/nm$^2$, dashed line and circles:
   $\sigma=-0.1$ e/nm$^2$. Symbols in all cases stand for numerical
   solution, lines represent first order perturbation theory (in $\epsilon$).}
\end{center}
\end{figure}

In Figs.~\ref{fig3a} and~\ref{fig3} we depict the current-voltage
(I-U)-characteristics of the ion transport.
The comparison of analytical and numerical results show that the
currents for both the ``short pore'' (Fig.~\ref{fig3a}a) and the
``long pore'' (Fig.~\ref{fig3a}b) are well predicted by the first
order perturbation theory, cf. Eq.~(\ref{cur_tot_first}), in the
range of $U$ from $-0.1$ V to $0.1$ V, for both surface charge
densities $\sigma=-0.02$ e/nm$^2$ and $\sigma=-0.1$ e/nm$^2$. 
However, for the larger density, the agreement deteriorates
for the ``short pore''. As expected, the discrepancy grows further
with increasing the absolute value of the applied voltage $U$ across
the nanopore, cf. Fig.~\ref{fig3}.

%In Fig.~\ref{fig4}, we compare the numerical and analytical results
%obtained with the proper boundary layers of finite width considered
%throughout this work and the numerical solutions obtained by imposing
%the Donnan potential jumps precisely at the channel boundaries in
%accordance with Eqs. (17)-(21) in Ref. \cite{Cervera} -- in that case
%the boundary layer width shrinks to zero.  Namely, the values of the
%Donnan potential jumps at the boundaries were extrapolated from the
%interior solution extended to the boundary.  They practically coincide
%\cite{Haenggi} in the considered $1/\lambda\to 0$ limit with ones
%following from Eqs. (17)-(21) of Ref. \cite{Cervera}.  Since the
%parameter $1/\lambda$ is very small even for the shorter channel
%considered, one might expect that the approximation of the Donnan
%potential jumps will work reasonably \cite{Cervera}.  The agreement is
%indeed excellent,  even  for the shorter channel, cf. Fig. 9.

The above examples evidence that the first order
perturbation expansion works well only for a relatively small charge
density $\sigma$ and not too large applied voltages. However,
in a wider range of voltages and
for larger charge densities one has to use a
higher order perturbation theory in $\epsilon$.

\begin{figure}[htb!]
\begin{center}
 \includegraphics[width=8.0cm]{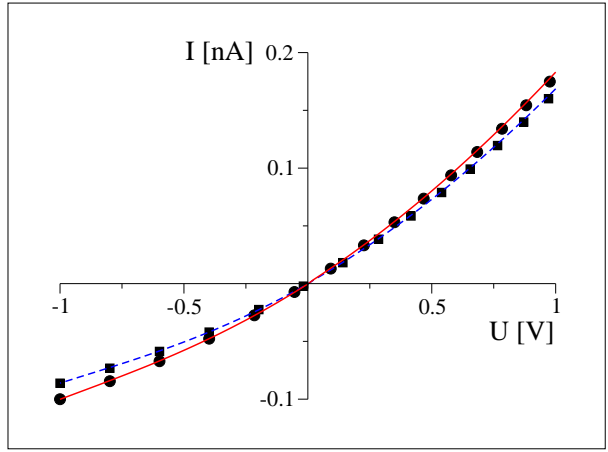}
 \caption{\label{fig6}
 (Color online) The potassium and chloride
currents in the
   ``short pore'' for $\sigma=-0.02$ e/nm$^2$.
The solid line represents the fourth order (in $\epsilon$)
perturbation
   calculation for potassium and the dashed line for the
   chloride current.  Symbols in all cases present the  numerical
   solution (circles: potassium current; squares: chloride)}
\end{center}
\end{figure}

In Fig.~\ref{fig6}, we depict the numerical $I-U$ current-voltage
characteristics for both the potassium $I_{\mathrm{K^+}}$, and the
chloride $I_{\mathrm{Cl^-}}$ currents versus the fourth order
perturbation theory results. The found agreement is rather good.
Undoubtedly, the characteristics are strongly non-linear and
asymmetric. Consequently the nanopore exhibits rectification
properties. We can analytically describe this effect in the first
order of $\epsilon$. Then, the individual ionic currents are
approximated by the following expansions: $I_{\mathrm{K^+}} = (I_0 -
J_0)/2 - \epsilon (I_1 - J_1)/2 + O(\epsilon^2)$ and
$I_{\mathrm{Cl^-}} = (I_0 + J_0)/2 - \epsilon (I_1 + J_1)/2 +
O(\epsilon^2)$. Interestingly, for both the potassium and the
chloride currents the difference between the absolute values of
positive and negative current branches is the same and equals to:
\begin{eqnarray}
\displaystyle |I_{\{\mathrm{K^+} ,\mathrm{Cl^-}\}}(U)| - |I_{\{\mathrm{K^+} ,\mathrm{Cl^-}\}}(-U)|  \approx
 -\epsilon I_1
=  -\frac{1}{6}D\pi\sigma\frac{R(L) - R(0)}{L}\bigg(\frac{eU}{k_{\rm{B}}T}\bigg)^2.
\end{eqnarray}
In a negatively charged nanopore, cations experience a potential
well, whereas anions sense a potential barrier. Thus, the potassium
concentration and also the potassium current  are always larger than
those of chloride, see Fig.~\ref{fig6}. This effect becomes enhanced
upon increasing the absolute value of the (negative) charge density
$\sigma$. However, the equivalent asymmetry of the $I-U$ dependence for individual currents
indicates that the rectification quality of the pore is independent of the ion sign
within this order of perturbation theory.
%To quantify the rectification effect, we define the
%quantity
%\begin{equation}
%\label{rectification}
% \displaystyle \alpha = (|I(U)| -
%|I(-U)|) / (|I(U)| + |I(-U)|)
%\end{equation}
%as a rectification-measure which is independent on the pore resistance.

\begin{figure}[htb!]
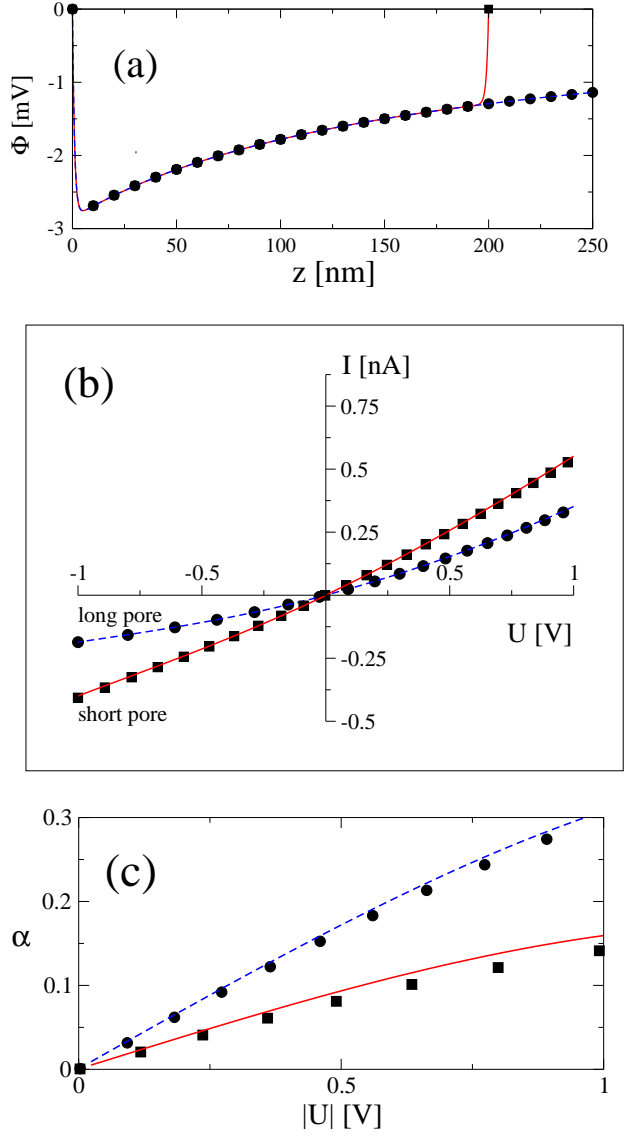

\begin{center}
\includegraphics[width=8.0cm]{pot_l_k}\\
\vspace{0.4cm}
\hspace{0.4cm}\includegraphics[width=8.0cm]{prad_l_k_bis}\\
\vspace{0.4cm}
\includegraphics[width=8.0cm]{alpha_l_k}
\end{center}
 \caption{\label{fig5}
(Color online) In the panel (a) we depict the potential profile
$\Phi(z)$ at
   equilibrium ($c_L=c_R=0.1$ M, $\Phi(0)=\Phi(L)=0$) for
   $\sigma=-0.02$ e/nm$^2$. In the panel (b) we depict the $I-U$ dependence
   and in panel (c) we show the
   rectification measure $\alpha$. The calculations are done for
   the ``long pore'': dashed line and circles; and the ``short   pore'': solid
   line and squares. The symbols in all cases stand for numerical
   solution; the
   lines represent the perturbation theory (in $\epsilon$): up to  second order (a),
   and in (b) and (c) up to fourth order.}
\end{figure}

Furthermore, in Fig.~\ref{fig5}a we compare the
electric potential profiles $\Phi(z)$ of the ``short pore'' (squares)
($L=200$ nm) and the ``long pore'' ($L=12000$ nm) (circles)
for the equilibrium situation.  They coincide up to
$z=195$ nm, because the angle of the cones of both nanopores are the
same. This implies that the electric field profiles in the narrow
parts of both channels are identical. Since the rectification
efficiency seemingly is  caused mostly by the sharp potential
variations in the narrow regions, one might naively expect that the
rectification quality will be similar for both cases, even if the
absolute $I-U$ characteristics for these two nanopores are rather
different, cf. Fig.~\ref{fig5}b. However, this would be an incorrect conclusion.

Indeed, the absolute
values of the \emph{total} electric current
$I = I_{\mathrm{K^+}}+I_{\mathrm{Cl^-}}$
is lower for the
``long pore'' as compared  to the ``short pore''. This result
corroborates with the higher resistance of the longer pore.
To quantify the rectification effect, we define the
quantity
\begin{equation}
\label{rectification}
 \displaystyle \alpha = \frac{|I(U)| -
|I(-U)|}{|I(U)| + |I(-U)|}
\end{equation}
as a rectification-measure. For it, the
lowest order perturbation theory result in Eq.~(\ref{cur_tot_first}) yields
\begin{eqnarray}
\alpha \approx -\frac{1}{24}\frac{1}{c_{\mathrm{bulk}} F}
  \left(\frac{2\sigma}{R(0)}-\frac{2\sigma}{R(L)} \right)\frac{e|U|}{k_{\rm{B}}T}.
\label{cur_tot_relative}
\end{eqnarray}
As seen from this expression, the quality of rectification $\alpha$,
does not depend on the channel length $L$
and is determined rather
by the difference of the inverse radii and by the surface charge.
Note that, in the above--described model, the volume density is given by
$\sigma(z) = 2\sigma/R(z)$ (Eq.~\ref{Poisson_mod}),
where $R(z)$ denotes a variable cone radius. The geometry
of the charged conical pore results not only in the presence of entropic potential
in the 1D current equation (Eq.~\ref{flux_steady}) but also in the asymmetrical
volume charge distribution $\sigma(z)$ in the reduced 1D Poisson equation (Eq.~\ref{Poisson_mod}).
This fact gives strong indication that in conical pores
the rectification is mainly due to different potential jumps
in the boundary layers at both ends of the channel.
This is in line with the previous statement that the $I-U$
characteristics of the nanopore depends crucially on the total
asymmetry of the potential profile \cite{JCP}.
Accordingly,
$\alpha$ reveals, cf. Fig.~\ref{fig5}c,  that the ``long
pore'' acts as a better rectifier, despite its larger resistance.
However, this is not due to a larger length, but due to a larger $R(L)$.

\begin{figure}[htb!]
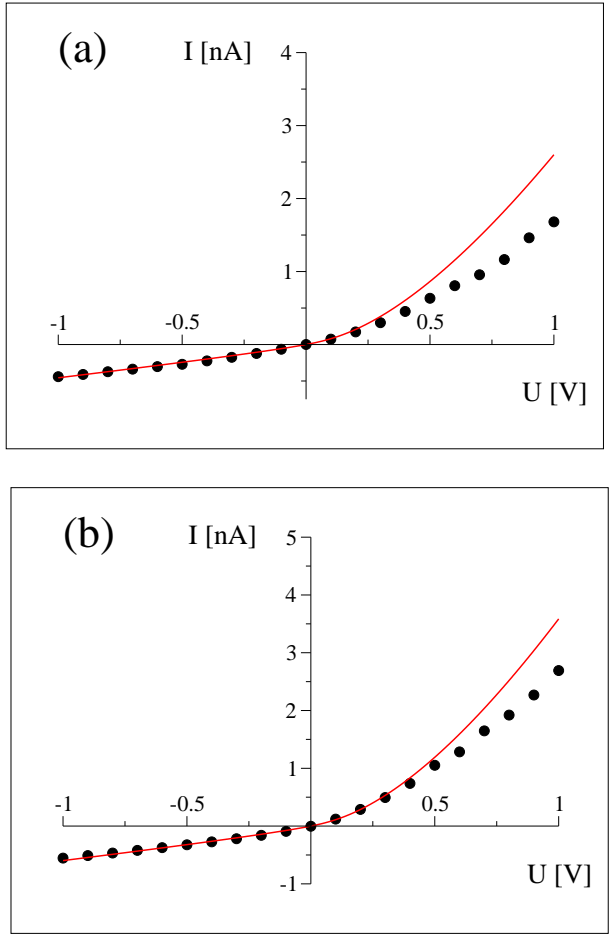

\begin{center}
 \includegraphics[width=8.0cm]{prad_SI_6_530_bis}
\vspace{0.4cm}

 \includegraphics[width=8.0cm]{prad_SI_7_680_bis}
 \caption{\label{fig7}
 (Color online) The current voltage ($I-U$) dependence in the
   experiment (symbols) and the 1D-PNP theory (lines).
   Only the
   numerical results are  given because the perturbation approach
   fails for the large surface charges used in the experiment.  We
   depict
   results for one value of the surface charge density
   $\sigma=-1.0 $ e/nm$^2$ and for $c_L=c_R=0.1$ M
   which lies within the range of
   experimentally measured values. We use two different pore geometries:
   panel (a) $R(z=0)=3$ nm, $R(L)=265$ nm, $L=12000$
   nm;
   panel (b) $R(z=0)=3.5$ nm, $R(L)=340$
   nm, $L=12000$ nm.}
\end{center}
\end{figure}

\subsubsection*{Theory {\it vs.} experiment}

In Fig.~\ref{fig7}, a comparison between the experimental data \cite{Siwy4}
and theoretical results is presented for two different nanopores of length $L=12 000$ nm
with the radii $R(0) = 3.5$ nm, $R(L) = 340$ nm and $R(0) = 3$ nm, $R(L)=265$ nm,
respectively. The experimental charge density is not known
precisely. The assumed value is around $\sigma=-1.0$ e/nm$^2$,
which corresponds to  $\epsilon \approx 6$.
This parameter
value is clearly beyond the validity range of
perturbation theory. Nevertheless, one can use numerics. Given the
approximate, reduced character of the studied
1D-Poisson-Nernst-Planck model, the semi-quantitative agreement
between the theory and the experiment is
quite satisfactory.

%Notably, a discrepancy occurs for large
%currents, where the prominent condition of a local equilibrium in
%the transverse direction is {\bf could be} violated. This breakdown of the 1D
%description {\bf seems to be} is similar to one observed for diffusive transport of
%neutral, uncharged particles \cite{Reguera1,Reguera2, Bezrukov}. A numerical
%study of the full three-dimensional set-up is clearly expected to
%improve  agreement.

%\subsubsection*{Local equilibrium in the transverse direction}

{\revision Indeed, the agreement is rather satisfactory for not too large
currents, $I < 0.5$ nA.} The simplest explanation for the
discrepancy occurring  for strong currents could be the violation of
the  condition of a local equilibrium in the transverse direction
(see Sec. One-dimensional model reduction). Then, this breakdown of
the 1D description would be similar in nature to the one observed
for diffusive transport of biased, non-interacting particles
\cite{Reguera1,Reguera2, Bezrukov}. To fully justify this statement,
one can follow the analysis in Ref.~\cite{Reguera2} to find
the different time scales characterizing  the problem, i.e. the time
scale for diffusion in the transverse direction and the time scale
associated with the drift of the $i$-th ion along the axial
direction. By comparison of this two time scales one could 
find out whether transported ions have
enough time to relax in the transversal direction. However, in the
present, more complicated situation the proposed consideration
results  in a very rough {\revision and somewhat vague estimation (because
of  a heterogeneous distribution of the ions and the corresponding
electrical field along the pore).} For better agreement a 3-D PNP
modeling would be called for.
%estimation.

%\subsubsection*{3D PNP modeling}

Such a 3D PNP study, would in natural way, reveal the dependence of
the electric potential on the radial coordinate $r$, and, in turn, a
heterogeneous distribution of ions in the transverse direction{\revision,
contrary to the model assumptions in Eq.~(\ref{constant}). Clearly,
the larger is the surface charge density, the stronger will be
the violation of these model assumptions.} Because of a negatively
charged channel wall, the positive ions should adhere to {\revision it}
forming an electrical double-layer {\revision of the width
$\xi_{\mathrm{D}}$ with an exponentially enhanced concentration of
cations. A physical criterion to disregard this complexity can be
obtained by demanding that the electrostatic energy per ion near the
channel wall (in the absence of external voltage), $\mathrm{e}V\sim
\mathrm{e}\sigma\xi_{\mathrm{D}}/(\epsilon_0\epsilon_w)$, is smaller
than $k_{\mathrm{B}}T$. For $\epsilon_w = 80$ and $\xi_{\mathrm{D}}=1$
nm this yields $|\sigma| < 0.1$ e/nm$^2$. For higher surface charge
densities (like in the experiment) not only the homogeneity
assumptions become increasingly violated, but also the surface
current contributions \cite{Chu,Chu2} cannot be neglected. These
effects are  beyond our description within a  1D-PNP reduction.
Moreover, for higher concentrations $c\sim 1$ M the ion-size effects
become important \cite{Jackson,Kilic} implying the presence of ion
correlations. This goes beyond the mean field description.}
%Successively, this part of the positive
%carriers would be excluded from transport.
%It might warrant a lower value of the observed currents in
%the positive branch of $I-U$ curve. On the other hand, the described process
%may also affect the surface current {\bf \cite{Chu,Chu2}.
%Surface current could become important when the surface
%charge density exceed $\frac{k_{\mathrm{B}}T\, \epsilon\,
%  \epsilon_{0}}{\mathrm{e} \, \xi_{\mathrm{D}}}$, that for our case is
%about $0.1$ e/nm$^{2}$, due to formation of the Stern layer of condensed ions.
%However, the treatment of surface current contributions is
%  completely beyond our modelling.}

To summarize, a numerical study of the full three-dimensional set-up
would be  helpful in order to distinguish between the principal
shortcomings of the PNP approach {\revision and the failure of our 1D
reduction for large currents and/or high surface charge densities.
In particular, the overall failure of the PNP approach can be due to
formation of the Stern layer of condensed counter-ions on the pore
wall \cite{Jackson}. Such a layer will screen and renormalize (i.e.
decrease) the surface charge density ``seen'' by other mobile ions,
thus reducing  the ``volume'' rectification effect. A proper treatment of the surface current rectification
effects in such a layer would bring about the theme of strong
correlations in the ionic transport which is  beyond the scope of
this work.}
%as such, a failure of the described 1D reduction because of the equilibration assumption
%which cannot be justified for sufficiently strong currents.

\section{Summary and Conclusion}

In this paper we provided an analytical treatment  of the problem of
current rectification by artificial conical nanopores made in
synthetic membranes of about submicron-to-micron width.  Within a
one-dimensional reduction of the original three-dimensional problem,
we have derived 1D-PNP equations, cf. Eqs. (12)-(14), which incorporate
both the entropic effects and the charge density renormalization due
to the variable diameter of the tube. These equations
generalize the 1D-PNP equations in Ref. \cite{Gardner} to the case
of pores of a variable diameter.

Furthermore, we provided a singular
perturbation treatment of the problem of ion conductance and
rectification within the reduced 1D-PNP description 
implementing rigorous boundary conditions (\ref{boundary}).
Our theory applies to nanopores with lengths exceeding largely
the Debye length and the pore diameter.
The developed theory corresponds
precisely to the experiments done by Siwy et al. \cite{Siwy4}.
The validity range of perturbation theory requires, however, that
the channel wall is charged weakly.
Then, it agrees well with the numerical solution of the 1D-PNP
problem for charge densities up to $|\sigma| = 0.02$ e/nm$^2$.
Unfortunately,
the experimental charge densities are much larger, about
$|\sigma|\sim 1$ e/nm$^2$. Here the perturbation theory fails. 
However, the 1D-PNP equations can be integrated numerically
and the obtained numerical solutions
provide a good agreement with the experimental
data by Siwy et al. \cite{Siwy4} for sufficiently small currents.

Moreover, we quantify the rectification properties of conical nanopores.
In the lowest order perturbation theory
we obtained analytic formulas for the rectification-current $I$ and
the rectification measure $\alpha$, see Eq.~(\ref{rectification}).
The latter measure clearly indicates
that the rectification property
is caused by the difference in the volume charge density
of the fixed charges.
In other words, the rectification is due to an asymmetry in the
potential jumps at the channel ends.

A discrepancy between the experiment and the 1D-PNP modeling
for large positive voltages most
likely indicates the violation of the condition of local
equilibrium in the transverse direction at strong currents, cf. Fig.~\ref{fig7}.
For this reason, the description within the 1D-PNP modeling 
breaks down and a full 3D-PNP treatment becomes necessary. 
We conjecture that the numerical solution of the 3D-PNP description
will provide satisfactory agreement with the experimental data for
the electrolytic solutions of mono-valence ions.

\acknowledgments

The authors express thanks to Prof.  Zuzanna Siwy for providing us
with her experimental data and for helpful discussion. This work has
been supported by Volkswagen Foundation (project number I/80424),
the Alexander von Humboldt Foundation (I. D. K.), the DFG (research
center, SFB-486, project A10), and by the German Excellence
Initiative via the \textit {Nanosystems Initiative Munich
(NIM)} (P.H.).
\appendix*

\section{Study case of Singular Perturbation Theory }

We consider a set of two coupled differential equations
(\ref{current_dim}), (\ref{Poisson_dim}) with the boundary
conditions given in Eq.~(\ref{boundary}). Each of these equations one
can be presented  in the form
\begin{equation}
\displaystyle P_{1/\lambda, \epsilon}(y) = 0
\end{equation}
where $\displaystyle y = \left\{ \bar{c}_{\mathrm{K^+}}, \bar{c}_{\mathrm{Cl^-}}, \Phi
\right\}$ and where two   small parameters $1/\lambda$ and
$\epsilon$ determine the behavior of the solution. The solution
shall be denoted by $\displaystyle y^{(1/\lambda)}_{\epsilon}$. The
limiting problem ($1/\lambda=0$)
\begin{equation}
P_{0, \epsilon}(y) = 0
\end{equation}
possesses  the solution $y^{(0)}_{\epsilon}$. The character of the
problem changes discontinuously at $1/\lambda = 0$. This implies
that we have a singular perturbation in this very parameter.

\subsubsection*{The Outer Approximation}
In the outer region (corresponding to the interior of
  the 1-D channel) we approximate $y^{(0)}_{\rm{outer}}(z) = \left\{
    \bar{c}^{(0)}_{\mathrm{K^+}}(z), \bar{c}^{(0)}_{\mathrm{Cl^-}}(z),
    \Phi^{(0)}(z) \right\}$
  by use of the regular perturbation expansion:
%\begin{equation}
\begin{align}
y^{(0)}_{\rm{outer}}(\epsilon,z) = y_{0}(z) + \epsilon y_{1}(z) +
\epsilon^2 y_{2}(z) + \dots .
\end{align}
%\end{equation}

\subsubsection*{The Boundary Layers}
The region near the boundaries at $z = 0$ and $z=1$ wherein
  $y$ is changing rapidly presents the boundary layer (left and
  right, respectively). The outer expansion looses validity there.  We
  re-scale the problem near $z = 0$ ($z=1$) by setting
  $\displaystyle \zeta = \lambda z$ ($\displaystyle \chi = \lambda
  (z-1)$), and express the various functions $\left\{
    \bar{c}^{(0)}_{\mathrm{K^+}}(z), \bar{c}^{(0)}_{\mathrm{Cl^-}}(z),
    \Phi^{(0)}(z) \right\}$
  in terms of the new coordinates as $\left\{ p^L(\zeta;1/\lambda),
    n^L(\zeta;1/\lambda), \phi^L(\zeta;1/\lambda) \right\}$,
    and $\left\{
    p^R(\chi;1/\lambda), n^R(\chi;1/\lambda),
    \phi^R(\chi;1/\lambda) \right\}$, in the left and right
    boundary layers, respectively. Next, considering
  the limit expansions obtained by holding $\zeta$ and $\chi$ fixed
  and letting $1/\lambda\to 0$, the problem can be solved in terms
  of the
 regular expansion $y^{(0)}_{L}(\epsilon, z)$ and $y^{(0)}_{R}(\epsilon, z)$.

\subsubsection*{Matching procedure}
We choose four constants which allow that  $y_{L}(\epsilon,z)$
  and $y_{\rm{outer}}(\epsilon, z)$, $y_{R}(\epsilon, z)$ and
  $y_{\rm{outer}}(\epsilon, z)$, coincide for each order of the expansion in
  powers of $\epsilon$ (as $1/\lambda\to 0$) in some intermediate zone
  between the left boundary layer and the outer region, and the right
  boundary layer and the outer region, respectively. This yields:

%\end{multline}
%\begin{equation}
\begin{eqnarray}
\vspace{10pt}
\displaystyle J_{0} = & 0, &
\displaystyle C_{0} =  2\pi c,\nonumber \\
\displaystyle I_{0} = & -C_{0}(1+\gamma)\Phi_R, &
\displaystyle E_{0} =  \frac{(1+\gamma)}{\gamma}\Phi_R. \nonumber \\
\vspace{10pt}
\displaystyle J_{1} = &
%-\frac{\pi}{2}\frac{ I_{0}}{{C}_{0}}
%\frac{\gamma+2}{\gamma+1} =
\displaystyle\frac{1}{2}(2 + \gamma)\pi\Phi_R, &
\displaystyle C_{1} = 
%-\frac{\pi}{2}
%\frac{ I_{0}}{{C}_{0}}\frac{1}{\gamma(\gamma+1)} =
\frac{\pi}{2\gamma}\Phi_R,\nonumber \\
%\end{array}
%\end{equation}
%and
%\begin{equation}
%\begin{array}{c}
%\vspace{10pt}
%\begin{multline}
\displaystyle I_{1} = &
%\frac{(1+\gamma)}{\gamma}\frac{I_{0}}{C_{0}}
%\bigg\{
%\frac{I_{0}}{6\gamma C_{0}}
%\left(\frac{1}{(1+\gamma)^3}-1\right)
%\\ + \frac{J_{1}}{2\gamma}
%\left(\frac{1}{(1+\gamma)^2}-1\right)
%+C_{1}\frac{\gamma}{1+\gamma}\bigg\},
\displaystyle\frac{1}{12}\pi\gamma\Phi^2_R, &
\displaystyle E_{1} = 
%-\frac{1}{2c_{L}} - \frac{I_{0}}{C^{2}_{0}\gamma}
%\bigg(
%\frac{I_{0}}{6\gamma C_{0}}+
%\frac{J_{1}}{2\gamma}
%-C_{1}\bigg)\\ - \frac{\left(I_{1} -
%\gamma\right)}{C_{0}\gamma}.
- \frac{(2 + \gamma)\pi}{12C_0\gamma^2}\Phi^2_R.
%\end{multline}
%\end{equation}
\end{eqnarray}
%\end{equation}

To obtain an approximate solution $y_u$ that is valid {\it
uniformly} on $[0, 1]$ we finally add the boundary and outer
approximations and subtract their common limit in the intermediate
zone.

\end{document}